\documentclass[twocolumn]{article}
\usepackage[margin=0.7in]{geometry}

\usepackage[font={footnotesize,sf}]{caption} % Smaller caption font
\usepackage{graphicx, subfigure, authblk, svg} % Required for inserting images
\usepackage{amssymb, amsmath} 
\usepackage[switch]{lineno} % line number
\usepackage{cuted} % For widetext environment
\DeclareCaptionFormat{lined}{\hrulefill\par#1#2#3\par\hrulefill}
\usepackage[style=numeric, sorting=none]{biblatex}
\title{Eukaryotes evade information storage-replication rate trade-off with endosymbiont assistance leading to larger genomes}
\author{Parthasarathi Sahu}
\author{Sashikanta Barik}
\author{Koushik Ghosh}
\author{Hemachander Subramanian\thanks{Corresponding author: hsubramanian.phy@nitdgp.ac.in}}
\affil{Department of Physics, National Institute of Technology Durgapur, India}
\date{}

\begin{document}

\maketitle

%\linenumbers

\begin{strip}
\hrule

\begin{abstract}

Genome length varies widely among organisms, from compact genomes of prokaryotes to vast and complex genomes of eukaryotes. In this study, we theoretically identify the evolutionary pressures that may have driven this divergence in genome length. We use a parameter-free model to study genome length evolution under selection pressure to minimize replication time and maximize information storage capacity. We show that prokaryotes tend to reduce genome length, constrained by a single replication origin, while eukaryotes expand their genomes by incorporating multiple replication origins. We propose a connection between genome length and cellular energetics, suggesting that endosymbiotic organelles, mitochondria and chloroplasts, evolutionarily regulate the number of replication origins, thereby influencing genome length in eukaryotes. We show that the above two selection pressures also lead to strict equalization of the number of purines and their corresponding base-pairing pyrimidines within a single DNA strand, known as Chagraff's second parity rule, a hitherto unexplained observation in genomes of nearly all known species. This arises from the symmetrization of replichore length, another observation that has been shown to hold across species, which our model reproduces. The model also reproduces other experimentally observed phenomena, such as a general preference for deletions over insertions, and elongation and high variance of genome lengths under reduced selection pressure for replication rate, termed the C-value paradox. We highlight the possibility of regulation of the firing of latent replication origins in response to cues from the extracellular environment leading to the regulation of cell cycle rates in multicellular eukaryotes.

\end{abstract}

\vspace{0.5\baselineskip}
\noindent \textbf{Keywords:} \textit{genome size evolution, prokaryote-eukaryote divergence, Chargaff's second parity rule, C-value paradox, endosymbiosis, latent origins, GC-Skew, origins of replication, replichores}

\vspace{0.5\baselineskip}
\hrule
\end{strip}

\section*{Significance Statement}
Understanding the forces shaping genome architecture is a long-standing challenge in evolutionary biology. Our study demonstrates that the balance between replication speed and information storage, constrained by cellular energetics, drives the divergence in genome lengths between prokaryotes and eukaryotes. By quantifying selection pressure as the ratio of replication time to genomic information content, we show that this pressure enforces adaptive constraints, giving rise to features such as symmetric replichores and Chargaff’s second parity rule. These insights not only help us resolve an enduring evolutionary puzzle, but also offer a unified framework linking genome organization, cellular specialization, and even potential mechanisms underlying carcinogenesis.

\section*{Introduction}
\label{introduction}

Life on Earth began approximately 3.7 billion years ago and evolved from simpler forms to complex and diverse organisms observed today, shaped by various selection pressures \cite{gregory2009understanding, dyson1999origins}. Organisms are broadly classified into two groups: prokaryotes and eukaryotes. Prokaryotes are characterized by simpler structures, whereas eukaryotes are generally more complex and have evolved from prokaryotes, with defining characteristics such as endosymbiotic relationships, nuclear membranes, huge variance in genome size, etc. \cite{gupta1996origin, embley2006eukaryotic, martin2015endosymbiotic}. Despite emerging earlier in Earth’s history, prokaryotes maintain smaller genomes and show less morphological evolution compared to eukaryotes \cite{sela2016theory, lynch2003origins, rocha2008organization}. The constraints limiting the complex morphological evolution of prokaryotes are debated \cite{lane2011energetics, chiyomaru2020revisiting, mira2001deletional}. 

%%%% New
The tendency of prokaryotes to acquire compact genomes is extensively modeled, with models constructed to include impacts of population size, environmental perturbations, and selection for metabolic efficiency under nutrient limitation \cite{sela2016theory, bentkowski2015model, martinez2022genome, rodriguez2023linking}. The evolutionary forces shaping eukaryotic genome length remain comparatively under-explored. In eukaryotes, current frameworks have largely focused on the impact of mutational mechanisms \cite{fischer2014model, colnaghi2020genome} and energetic constraints on genome length \cite{lane2010energetics}.

% One of the major differences between these two groups is genome length, with prokaryotes having a 20-fold variation in genome length, whereas eukaryotic genomes vary 200,000 fold \cite{hou2009distinct, graur1997molecular}. It is also evident that this dramatic spread in eukaryotic genome length doesn't correspond to organismal complexity \cite{markov2010relationship}. The lack of correlation of genome length with organismal complexity is termed as `C-value paradox' \cite{thomas1971genetic}. There are two classes of explanation for this paradox: (a) mutational pressure theories \cite{ohno1972so, pagel1992variation} and (b) optimal DNA theories \cite{gregory2001coincidence, cavalier1982skeletal}. Under the first class,  junk and selfish DNA are considered responsible for the expansion of genome length in eukaryotes. It is argued that the inevitable expansion of efficient self-replicators would produce a constant pressure to increase genome length, and this expansion is curtailed only when the additional genetic material becomes excessively burdensome for the host cell \cite{szathmary1995major}. In the second class, phenotypic constraints, such as cell volume, evolutionarily select a specific genome length. 

Despite extensive studies, a simple explanation for such a dramatic divergence of genome length between prokaryotes and eukaryotes is lacking. In this study, we use a very simple, parameter-free model that incorporates the influence of two primary evolutionary forces: faster replication and enhanced information storage capacity, to study the evolution of genome length across these two domains of life. We show that the genome lengths of prokaryotes and eukaryotes diverge under the same selection pressure, if we restrict prokaryotes to have a single replication origin, while allowing eukaryotes to have multiple origins. We argue that this differentiation stems from access to the energy supply of mitochondria (or chloroplasts), as evidenced by the observed correlation between the number of mitochondria (or chloroplasts) and the length of the genome in eukaryotes. Surprisingly, the model also reproduces multiple other observations that hold for nearly all species, such as the equality of purines and pyrimidines on a single strand of DNA, called Chagraff's second parity rule (PR-2), replichore length symmetrization (see below), a preference for deletions over insertion mutations, and a huge variance in the genome lengths seen in eukaryotes, called the C-Value paradox.   

\section*{The Model}
\label{model}
To study the effect of the aforementioned selection pressures on genome length, i.e. faster replication and higher information storage capacity, we utilize a model, where a pool of $N$ identical sequences is evolved over $m$ generations under these selection pressures. The initial sequence pool consists of $N$ identical sequences, containing purines or pyrimidines, homogeneously across the sequence, with all purines on one strand and all pyrimidines on the complementary strand (e.g., $5'$-RRRRRRR-$3'/3'$-YYYYYYY-$5'$). Each generation involves two major steps: (i) replication of $N$ sequences and mutation of all daughter sequences in the pool and (ii) selection of half of the sequences in the pool based on their ability to satisfy the above two selection pressures. These two steps are repeated $m$ times, and the time-evolution of the sequence length, averaged over all the $N$ sequences, is recorded for every generation, for subsequent analysis.

\begin{figure}[h!]
    \centering
    \includegraphics[width=\linewidth]{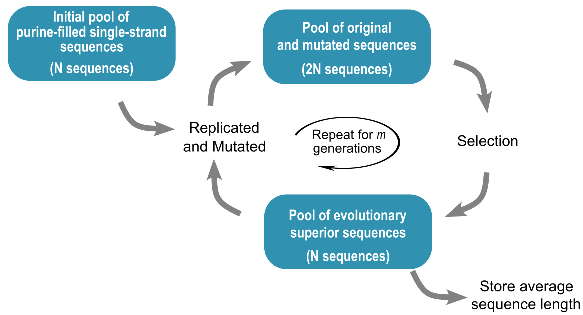}
    \caption{Algorithm of the model. Evaluation of the impact of the two selection pressures, fast replication and high information storage capacity, on the genome length of an organism. An initial pool of $N$ sequences is evolved over $m$ generations that involve two recurring steps: replication and mutation of all the sequences in the pool and applying selection pressure to extract the fittest sequences for the next generation. An initial pool of $N$ identical sequences, composed of all purines or all pyrimidines, are replicated, and the daughter sequences are mutated, producing a pool of $2N$ sequences. Selection acts on this pool, removing $N$ less-fit sequences that do not satisfy the selection pressure adequately. This replication-selection cycle is repeated $m$ times, and the average genome length at every generation is recorded.}
    \label{fig: algorithm}
\end{figure}

\begin{figure}[h!]
    \centering
    \includegraphics[width=\linewidth]{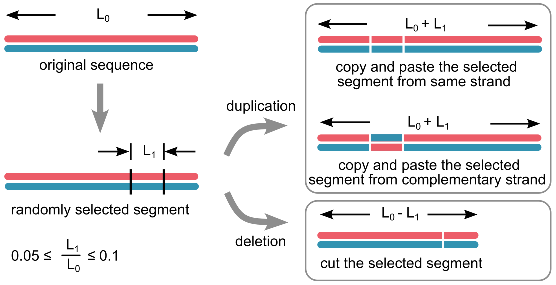}
    \caption{Mutation of a given genome. A DNA double strand, from the initial pool, with a homogeneous distribution of purines on one strand (red) and pyrimidines on its complement (blue). A mutation involves either a deletion or a duplication of a segment of a random length of 5-10\% of the total genome length, chosen at a random location of the daughter genome. The mutation results in either a decrease or an increase in the length of the genome by 5-10\%. The duplicated fragment can be from either strand, thereby altering the composition of purines/pyrimidines in each strand.}
    \label{fig: mutation}
\end{figure}

\noindent \textbf{Mutation:} We implement large-scale genomic mutations through random deletions or duplications of regions in the daughter sequence that comprise 5\% to 10\% of the total genome length (Fig. \ref{fig: mutation}). These large-scale mutations represent well-documented drivers of evolutionary processes \cite{sela2016theory, moran1958random, wolf2009universal, albrecht2007inversions, mira2001deletional}. The size and location of these mutations are chosen stochastically. During a mutation involving duplication, the duplicating fragment is randomly chosen from either the original strand or the complementary strand. This mechanism ensures that over generations, each strand can have varying amounts of both purines and pyrimidines, even though the initial pool sequences are composed entirely of either purines or pyrimidines. In each generation, every sequence is replicated, and the replicated sequence in the pool undergoes a single mutation, i.e., a duplication or deletion. Following this, the sequence pool is expanded to include $N$ replicated and mutated sequences along with the $N$ unmutated sequences of the previous generation, resulting in a total of $2N$ sequences.

\noindent \textbf{Selection:}
Following mutations, $N$ sequences are selected from the pool of $2N$ for the next generation based on their ability to satisfy the selection pressure. The selection pressure is quantified through a factor $\gamma$, defined as

\begin{equation}\label{selection factor}
    \gamma = \frac{\textit{replication time}}{\textit{information storage capacity}}  
\end{equation}

To quantify information storage capacity, we utilize the metric of total genome length, since it provides the upper limit for total information storage. Similarly, as a proxy for the replication time for the entire genome, we consider the length of the longest \textit{replichore}, where, replichores are defined as disjoint segments of the genome that replicate independently of each other. This substitution is based on the following considerations: (a) Replichores replicate simultaneously and independently, parallelizing the replication process \cite{subramanian2020evolutionary, sahu2024high}. Although the firing of multiple replication origins across the genome is generally not synchronized \cite{boos2019origin, patel2006dna}, we assume ideal conditions, where origins fire and replication progress across each replichore independently and simultaneously. (b) For simplicity, it is assumed here that the replication speed of a replichore is constant throughout the length of the replichore, although the speed depends on the sequence and the availability of activated nucleotides, in general \cite{wang2011replication}. This simplifying assumption enables us to use the replichore length as a replacement for the replication time of the whole genome, in our evaluation of $\gamma$. Therefore, with the above assumptions, we redefine equation (\ref{selection factor}) as 

\begin{equation}\label{selection factor modified}
    \gamma = \frac{\textit{length of longest replichore}}{\textit{length of full genome}}  
\end{equation}

In the selection process, the factor $\gamma$ is calculated for all the $2N$ (parent and mutated daughter) sequences in the pool, and the $N$ sequences with the lowest values of $\gamma$ are extracted as the fittest sequences and are carried forward to the next generation.

To identify replichores in a genome, we first locate replication origins and termini by analyzing base composition asymmetry or nucleotide skews, i.e., the excess of G over C and A over T on any given stretch of a single strand, across the genome. This local violation of Chagraff's second parity rule \cite{rudner1968separation} is regularly utilized to find the replication origins and termini, by equating the locations of peaks and valleys in the skew plot to the replication origins and termini. This is a widely used technique for \textit{in-silico} prediction of replication origins since 1996 \cite{sahu2024high, lobry1996asymmetric, grigoriev1998analyzing, tillier2000contributions, mclean1998base}. In this study, we have used the purine-pyrimidine (RY) cumulative skew, $W$, of the sequences \cite{zhang2004identification, sernova2008identification}, where $W_{RY}(n)$ is defined as
$W_{RY}(n) = \Sigma _{i=1}^n (\delta _{S(i),R} - \delta _{S(i),Y})$.
Here,$S$ is a genomic sequence of length $N$ bp, composed of four nucleotides, classified into two groups: $R = \{G, A\}$, $Y = \{C, T\}$, and $n = 1 \dots N$. We have taken 1000 purine-filled single strands of length 1024 bp each as the initial sequence pool. These sequences are allowed to mutate, by accrual or deletion of genome fragments, leading to local distortions in the skew in each generation. In order to avoid identifying small-scale skew variations (peaks and valleys smaller than a certain length scale) as origins or termini, since these are not identified as origins or termini by origin-finding algorithms \cite{sernova2008identification, frank2000oriloc, zhang1994z, gao2008ori}, we concentrate only on large-scale skew variations and ignore small-scale ones. We use wavelet transforms to filter out these origins and termini resulting from small-scale variations, by down-sampling the genome sequence of length $N$ to a length of $N/2^w$, where $w$ is the wavelet level. To ensure that mutating fragments are not smaller than the wavelet compression scale $w$ and thus go unnoticed by the selection pressure, we choose a $w$ such that the smallest mutating fragment is larger than the compression factor; i.e., we impose the condition $0.05N > 2^w$, where $N$ is the full genome length. It should be emphasized that the qualitative divergence of prokaryotic and eukaryotic genome lengths does not depend on the wavelet level used in this down-sampling procedure. Once the origins and termini of the replication are identified, the lengths of the replichores were measured as the distances between neighboring origins and termini, the largest of which is chosen for the calculation of $\gamma$. A detailed description of the methodology for identifying replication origins and measuring replichore lengths is provided in the supplementary material.

In our model, we chose the population $N = 1000$, the initial sequence length of 1024 bp ($L_0$), the number of generations $m = 1000$, and used a $4$-level wavelet transformation. We repeated the experiment 100 times to ensure statistical robustness.

Our model also includes an upper threshold for the number of replication origins allowed in a genome ($Ori_{max}$) to prevent an uncontrolled explosion in genome length. Genomes with a replication origin count greater than $Ori_{max}$ are eliminated during the selection process. For prokaryotes, $Ori_{max}$ is set to 1, while for eukaryotes, $Ori_{max}$ is set to a value much greater than 1 (50 and 100). Although we use parameters such as $Ori_{max}$, wavelet levels, and mutation size 5\% - 10\% for computational convenience, the model itself is free of intrinsic parameters, and the observed divergence between the prokaryotic and eukaryotic genome length is completely insensitive to the parameters listed above.

\section*{Results}
\label{result}

\begin{figure}[h!]
    \centering
    \includegraphics[width=\linewidth]{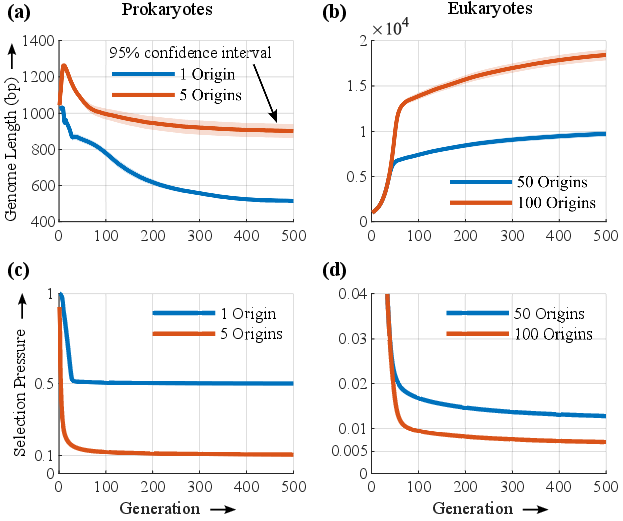}
    \caption{The evolution of genome length over generations. An initial population of 1000 purine-filled single-strand sequences is subjected to mutation and evolved under the selection pressure ($\gamma$) to minimize replication time while maximizing information storage capacity. The experiment is repeated 100 times, and the evolution of average genome length over 500 generations is shown, with a 95\% confidence interval. (a) When each genome in the pool is constrained to have a single replication origin, mimicking prokaryotic genomes, the average genome length decreases over generations. Also shown is the genome length evolution when a few more origins are allowed, as in the case of archaea. (b) In contrast, when the sequences are allowed to accommodate many more replication origins, mimicking eukaryotic genomes, the average genome length increases. This is because the expansion in genome length of eukaryotes does not come at the expense of replication time. Unlike prokaryotes, eukaryotes can parallelize replication across multiple replichores by replicating them independently and simultaneously, due to the presence of multiple origins, thereby reducing the replication time substantially, while maintaining a large genome length. Due to the constraint of single origin, prokaryotes cannot have more than two replichores, and hence cannot parallelize replication beyond these two, thus restricting their genome length. (c) and (d) Minimization of mutation pressure $\gamma$. For both prokaryotes and eukaryotes, $\gamma$ converges to $1/\text{number of replichores}$. This convergence implies symmetrization of replichore lengths across a genome. The initial pool has no origins, and hence one replichore, and the initial value of $\gamma$ is thus 1.}
    \label{fig: LenEvo}
\end{figure}

The variation in genome length (in bp) of prokaryotes and eukaryotes over $500$ generations is shown in Fig. \ref{fig: LenEvo}(a) and (b), respectively. The evolutionary minimization of the selection pressure $\gamma$, for prokaryotes and eukaryotes, is shown in Fig. \ref{fig: LenEvo}(c) and (d). 

% increase and stabilize
We observe that, in the absence of any restrictions on the number of replication origins, the genome tends to increase in length indefinitely (Fig. \ref{fig: LenEvo}b). However, taking into account the scarcity of monomer resources and energy required to replicate longer genomes, we restrict genomes to have a maximum of $Ori_{max}$ origins. When $Ori_{max}$ is set high ($>> 1$), the average genome length is observed to increase over generations. On the other hand, when $Ori_{max}$ is restricted to 1, mimicking a prokaryotic genome, the average length of the genome decreases over generations. Despite the applied selection pressure being identical in both scenarios, prokaryotic genomes tend to lose sequence length, while eukaryotes tend to elongate their genomes. This behavior mirrors the evolutionary divergence between the lengths of the prokaryotic and eukaryotic genomes. In the context of the model, the explanation for this pattern of genome evolution is as follows.

\begin{figure}[h!]
    \centering
    \includegraphics[width=\linewidth]{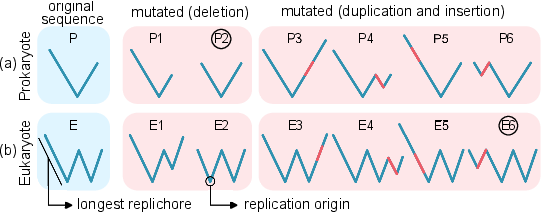}
    \caption{Effect of selection pressure on prokaryotic and eukaryotic genomes. The left panel, in blue background, shows the skew profiles of a prokaryotic and a eukaryotic genome of an intermediate generation, before mutation. The right panel, in red background, shows a few skew profiles of these genomes, after mutation. For eukaryotic genomes, mutations adding an extra replichore originating from a new replication origin (e.g., `E6') are preferred, as they increase information storage by elongating the genome without increasing the replication time (determined by the length of the longest replichore). In prokaryotic genomes, mutations that increase the length of the longest replichore (e.g., `P3' and `P5') or add a new replication origin (e.g., `P4' and `P6') are eliminated by selection pressure. Instead, mutations that shorten the longest replichore length, leading to a symmetrization of replichore lengths (e.g., `P2'), are favored, as they optimize the selection pressure ($\gamma_\text{opt} = 1/2$) by lowering the replication time.}
    \label{fig: selection}
\end{figure}

Consider a genome with an asymmetric purine-pyrimidine composition, where all nucleotides at the $5'-$ end of the replication origin are pyrimidines, and those at the $3'-$ end are purines. The cumulative skew profile of this sequence forms a ``V" shape, as illustrated in Fig. \ref{fig: selection}(a). Prokaryotic genomes with the two replichores emerging from their single origin exhibit such skew profiles. The single replication origin is located at the valley point of the ``V"-shaped skew profile \cite{grigoriev1998analyzing, tillier2000contributions, mclean1998base, gao2008ori}. The skew profile of a eukaryotic genome, on the other hand, has multiple concatenated ``V"-shapes, as shown in Fig. \ref{fig: selection}(b), where, each arm of each ``V" corresponds to a replichore, and the multiple valley points, to multiple origins. Mutations in the sequence alter this skew profile. The selection pressure $\gamma$ depends on whether the genome undergoes deletion or insertion, and which replichore arm (shorter or longer) is affected due to the mutation. Fig. \ref{fig: selection} shows a few skew profiles of the mutated sequences. Selection acts on the set of genomes with such altered skew profiles and prefers sequences that decrease the replication time and/or increase the information storage.

In the prokaryotic genome (P), mutations involving elongations are not preferred by the selection, since it either increases the longest replichore length (e.g. P3 and P5), resulting in an increase in overall replication time, or adds an extra valley point to the sequence (e.g. P4 and P6), which is eliminated by the selection pressure, as prokaryotes are restricted to have a single replication origin. However, a mutation that shortens the longest replichore (e.g., P2) is selected because it decreases the replication time. Unlike prokaryotic genomes, in eukaryotes, our selection algorithm allows for the addition of more origins, and hence more replichores (e.g. E4 and E6). If these new replichores are not the longest among all replichores, mutated sequences containing them will be selected due to their increased information storage capacity and neutral influence on replication time (e.g., E6). Therefore, the genome length of eukaryotes continues to increase over generations through the incorporation of new origins and hence replichores, until selection restricts further increase in the number of origins due to the upper limit $Ori_{max}$. 

In both prokaryotic and eukaryotic organisms, evolutionary selection pressures favor the symmetrization of replichore lengths. This phenomenon arises because replichores that are shorter than the longest replichore can undergo elongation through mutational processes, as this enhances the genome's information storage capacity without increasing its replication time, thereby reducing $\gamma$. This aligns with the general observation of symmetric replichore lengths in prokaryotes \cite{matthews2010fitness, darling2008dynamics, jespersen2024insertion} and the balanced distribution of short sequence motifs in eukaryotes \cite{albrecht2007inversions, prabhu1993symmetry}. Collectively, these selective forces drive two key outcomes: (1) the equilibration of replichore lengths and (2) the addition of one or more new origins, and hence replichores.

The observed genome length reduction in prokaryotes (Fig. \ref{fig: LenEvo}) is due to a bias of the selection pressure that favors deletions over insertions, although both mutational events are equally probable in our model. As explained in the previous paragraph, selection pressure favors the symmetrization of replichores. When two replichores of the prokaryote are of unequal length, symmetrization requires deletion of the longer replichore or insertion into the shorter replichore. Since our choice for the location of these two mutational events is entirely random and, therefore, evenly distributed over the lengths of the genome, the probability of choosing the shorter replichore for insertion or deletion will always be smaller than the probability of mutations occurring on the longer replichore. Although selection favors insertion into shorter replichore and deletion at longer replichore equally, since the latter is stochastically more favored, deletion occurs more frequently. This computational observation reproduces the strong evolutionary preference observed experimentally for deletions over insertions in prokaryotes \cite{kuo2009deletional, mira2001deletional, taylor2004occurrence, tao2007patterns, andersson2001pseudogenes, gregory2004insertion}.

An initial increase in the average genome length is seen in Fig. \ref{fig: LenEvo}. This is an artifact of our initial choice of genomes, each of which has been chosen as a homogeneous stretch of purines or pyrimidines. Therefore, these initial sequences have no origin and the length of the entire genome is equal to the length of a replichore, with $\gamma = 1$. The deletions in these sequences do not alter the value of $\gamma$, since they affect both the numerator and the denominator of $\gamma$ equally. However, duplications can reduce $\gamma$ when the duplicating fragment is from the complementary strand, introducing a new replichore and an origin. The reduction in $\gamma$ is due to the division of the genome into two replichores, due to the introduction of an origin, thereby reducing the replication time \cite{subramanian2020evolutionary}. As a result, within our model, during the early stages of evolution, deletions are not favored by selection, and the average genome length increases.

%-%-%-%-%-% New %-%-%-%-%-%
%-%-%-%-%-% New %-%-%-%-%-%

\subsubsection*{Symmetrization of replichore lengths leads to Chargaff’s Second Parity Rule}

The evolution of genomic sequences in our model begins with a pool of sequences composed entirely of purines (R) on the Watson-strand and complementary pyrimidines (Y) on the Crick-strand. In successive generations, mutation involving the exchange of strand fragments between complementary strands introduces strand heterogeneity, that is, sequences with interspersed R and Y bases. As demonstrated in the earlier section, evolution under the defined selection pressure $\gamma$ drives sequences to have replichores of equal length. Half of the replichores, positioned to the left of each origin, become pyrimidine-rich, while the remaining half, to the right of origins, are purine-rich. This symmetry in the cumulative nucleotide skew around replication origins \cite{lobry1996asymmetric} results in global parity in the purine and pyrimidine content throughout the genome, leading to Chargaff's second parity rule (PR-2).

Chargaff’s first parity rule, identified before the discovery of DNA's double-stranded structure \cite{chargaff1950chemical, chargaff1952composition}, revealed equal counts of adenine (A) and thymine (T), as well as guanine (G) and cytosine (C) in double-stranded DNA (dsDNA), a pattern now understood as a consequence of Watson-Crick base pairing. In contrast, Chargaff’s second parity rule (PR-2), which extends this symmetry to individual strands of dsDNA, lacks a universally accepted mechanistic explanation \cite{sueoka1995intrastrand, mitchell2006test, baisnee2002complementary, forsdyke2021neutralism}. Early hypotheses attributed PR-2 to adaptive intra-strand stem-loop formation, favoring local sequence inversions to achieve the functional benefits of self-complementarity \cite{forsdyke1995relative, forsdyke2021neutralism}. However, this rationale fails to account for PR-2’s prevalence in non-coding regions, where selective pressures for secondary structures are weak \cite{zhang2010limited}. Alternative theories propose PR-2 as a manifestation of the law of large numbers or an emergent property of entropy maximization in large genomes, where stochastic shuffling of sequences via inversions and transpositions homogenizes base composition over time \cite{lobry1995properties, albrecht2006asymptotically, hart2012gibbs, fariselli2021dna, matkarimov2023chargaff, pflughaupt2023generalised}. However, these mechanisms rely on \textit{no strand-bias} assumptions and do not account for an asymmetry in base substitution frequencies between purine and pyrimidine, i.e. $R \to Y \neq Y \to R$ \cite{vartanian1996hypermutagenic, frank1999asymmetric}. More importantly, while these theories account for global compositional symmetries, they neither explain nor reproduce local nucleotide compositional skews around replication origins that are prevalent in genomes of nearly all species \cite{lobry1996asymmetric, grigoriev1998analyzing, tillier2000contributions, mclean1998base, bartholdy2015allele}. The near-universal presence of local violations of PR-2 points to their importance, specifically for replication origin functionality. Stochastic nucleotide shuffling would erase these local PR-2 violations as well, and hence will be counterproductive.

\begin{figure}[h!]
    \centering
    \includegraphics[width=.6\linewidth]{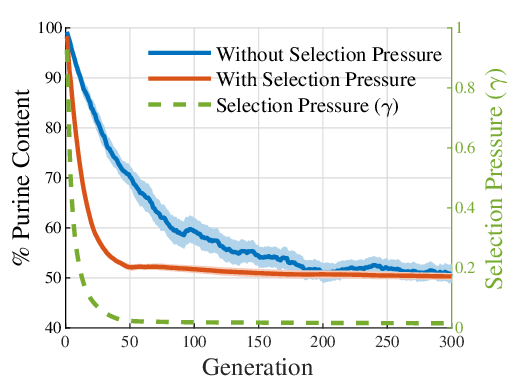}
    \caption{Evolutionary trajectories of base composition over generations. An initial pool of double-stranded DNA, with the Watson strand composed entirely of purines (R) and the Crick strand, of pyrimidines (Y), is evolved through mutations (inversions and inverted transpositions). The variation in purine content of the Watson strand over generations is shown, with shaded regions indicating 95\% confidence interval. \textbf{Blue curve:} Strand evolution without any selection pressure, exhibiting gradual R/Y parity in DNA single-strand, due to stochastic inter-strand shuffling. This demonstrates PR-2 emerging from stochasticity alone. These sequences do not exhibit any structured nucleotide skew profiles associated with replication origins, seen in most genomes. \textbf{Red curve:} Under selection, R/Y parity emerges more rapidly via replichore symmetrization, which is a direct consequence of the selection pressure to minimize replication time by balancing replichore lengths. Selection maintains the nucleotide skew profiles around replication origins, aligning with experimental observations. Selection also preserves PR-2 by eliminating mutational biases for specific nucleotides, removing the necessity for the no-strand bias assumption, used in explanations based on neutral processes. }
    \label{fig: WOpressure}
\end{figure}

In our framework, although we incorporate inter-strand shuffling, PR-2 emerges not as a passive outcome of random sequence shuffling but as an adaptive response to selection pressure (Fig. \ref{fig: WOpressure}). Here, selection pressure eliminates any bias in the base substitution frequencies between purines and pyrimidines. Any alteration in the symmetric, V-shaped profile of the cumulative nucleotide skew (see Fig. \ref{fig: selection}) due to the bias in base substitution frequencies is not tolerated by the selection, since it would adversely affect the replication time, by making the replichore lengths asymmetric. Therefore, selection acts against such biased substitutions, restoring the symmetry of the cumulative skew diagram and that of the replichore lengths, and hence the global equivalence in the purine-pyrimidine content of a DNA single-strand. The evolutionary trajectory of purine content in our simulations (Fig. \ref{fig: WOpressure}) illustrates this dynamic, demonstrating a rapid convergence to R/Y parity concurrent with replichore symmetrization. Note that, although the mean R/Y composition equilibrates to $50\%$ in both selective and neutral evolutionary processes (red and blue curves in Fig. \ref{fig: WOpressure}), the variance in R/Y composition is larger for neutral evolution, when compared with the evolution under selective pressure. This suggests that strong selective pressure leads to strict compliance with PR-2, whereas, non-adaptive evolution allows for deviations from PR-2. If the selection pressure for short replication time is weak or non-existent, as in the case of mitochondria, and for plasmids and viroids, the PR-2 compliance requirement vanishes, according to our model. There is a lack of need to maintain symmetric replichore lengths to minimize replication time in these genomes, since the rate-limiting step for their replication is the replication time of the host genomes, which are much larger. This prediction has been validated experimentally \cite{mitchell2006test, nikolaou2006deviations}. Chloroplasts' replication is independent of its host cell cycle \cite{kabeya2013chloroplast}, and it is, therefore, PR-2 compliant.

\subsubsection*{Influence of endosymbiont power supply on genome length}

\begin{figure}[h!]
    \centering
    \subfigure[]{
        \includegraphics[width=0.46\linewidth, trim= 5 0 19 0, clip]{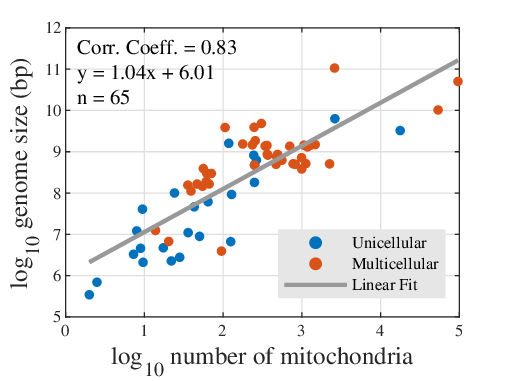}
        \label{fig:subfig1}
    }
    \subfigure[]{
        \includegraphics[width=0.46\linewidth, trim= 5 0 19 0, clip]{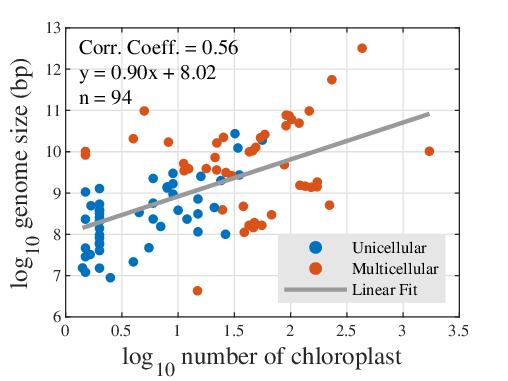}
        \label{fig:subfig2}
    }
    \caption{(a) Correlation between the number of mitochondria and genome length in $65$ eukaryotic cells. (b) Correlation between the number of chloroplasts and genome length in $94$ eukaryotic cells. These plots are produced using data from Jordan G. Okie et al. \cite{okie2016major}. We estimate the genome length from cell volume using the relation $V = kC^{\alpha}$ \cite{Gregory2000}, with $k = 3.04 \times 10^{-5}$ $\mu m^3/bp$ and $\alpha = 0.89$. The genome length is positively correlated with mitochondrial or chloroplast count, supporting the argument that the replication of longer genomes of eukaryotes is carried out with the energy provided by the endosymbionts.}
    \label{fig: MitoVsGenomeSize}
\end{figure}

According to our model above, prokaryotes tend to minimize their genome length due to the limitation of a single replication origin, whereas eukaryotes tend to acquire genome content, because of their ability to accommodate multiple replication origins. The rationale for our choice to restrict the number of origins of prokaryotes to one, while allowing the eukaryotes to have many more is to align the model with observations. A deeper reason for this choice lies in the bioenergetic requirement for maintaining multiple replication origins, which is to provide activated monomers and energy supply for the replication machinery \textit{simultaneously} to multiple replicating segments of a large genome. This requirement is met through the endosymbiotic relationship of eukaryotes with mitochondria or chloroplasts \cite{lane2010energetics, lane2011energetics}. Jordan G. Okie \textit{ et al.} have identified a linear relationship between mitochondrial and chloroplast count and cell volume in a number of eukaryotes \cite{okie2016major}. Since there is a well-known allometric relationship between cell volume and genome length \cite{gregory2001coincidence, Gregory2000, olmo1983nucleotype, mg1973dna}, we have converted the cell volume data of Jordan G. Okie \textit{et al.} to that of genome length and plotted the variation of genome length as a function of mitochondrial/chloroplast count. Fig. \ref{fig: MitoVsGenomeSize} shows this linear relationship between the above two variables. This suggests that the length of the eukaryotic genome is in part determined by the availability of the power supply to simultaneously replicate multiple genomic segments, provided by multiple mitochondria or chloroplasts.

This limitation imposed on genome length by cellular energetics is taken into account in our model by limiting the number of origins to a set value, $Ori_{max}$. In prokaryotes, which lack mitochondria/chloroplasts, the number of origins is generally limited to 1 (or a few), and therefore we take $Ori_{max} = 1$. In eukaryotes, the number of origins can be of the order of tens of thousands, and are limited only by power supply availability, as argued above (\ref{fig: MitoVsGenomeSize}). We therefore set $Ori_{max}$ to 50 or 100, to explore the genome length divergence between prokaryotes and eukaryotes.

\subsubsection*{Upper and lower bound of genome length}

As explained before, the selection pressure symmetrizes the replichore length and adds new origins, and thus replichores, to maximize information storage capacity. After evolving the sequences for a considerable number of generations, the selection produces genomes with nearly equal replichore lengths and the maximum number of origins ($Ori_{max}=1$ for prokaryotes and $Ori_{max}>>1$ for eukaryotes), thus minimizing the selection pressure. The theoretical optimum genome length under these conditions is 
\begin{equation} \label{eq: Optimum Length}
    \begin{aligned}
        L_{opt} &\approx \text{length of each replichore} \times \text{max no. of replichores} \\
                &= \text{length of each replichore} \times 2\: Ori_{max}
    \end{aligned}
\end{equation}
Therefore the optimum selection pressure is
\begin{equation}\label{eq: Optimum Gamma}
    \gamma _{opt} = 1/\text{number of replichores},
\end{equation}
The selection pressure can be seen to converge to this optimum in Fig. \ref{fig: LenEvo}(c) \& (d). For prokaryotes, $\gamma_{opt} = \frac{1}{2}$, since a single origin can only support two replichores. Whereas, for eukaryotes, $\gamma_{opt}$ is $0.01$ and $0.005$, since, at optimum genome length, they will have $\approx 100$ and $\approx 200$ replichores, corresponding to 50 and 100 origins, respectively.

\noindent \textbf{Lower bound:} The model introduces a wavelet-based resolution limit for distinguishing replication origins. Following a $w$-level wavelet transform of the genome, pairs of peaks or valleys in the nucleotide skew profile separated by fewer than $2^w$ nucleotides become indistinguishable, as the transform disregards small-scale variations. This imposes a minimum threshold of $2^w$ base pairs (bp) on replichore lengths. This threshold may serve as a proxy for the minimal genomic information necessary for cellular viability, reflecting evolutionary constraints on the smallest genome length.  Consequently, in our simulations, the smallest genome length achievable for prokaryotes is of the order of $2 \times 2^w$ nucleotides. Within the model, this lower bound reflects the computational resolution limit for identification of origins, whereas, in the evolutionary dynamics captured by the model, this lower bound reflects the need for sufficient informational complexity for cellular maintenance and replication machinery \cite{markov2010relationship, gonzalez1994smallest, glass2006essential, fraser1995minimal, gil2004determination}. Our choice of a specific wavelet level ensures that the genome length does not reduce below a certain viability limit.

\noindent \textbf{Upper bound:} One can make an interesting observation from the eq. \ref{eq: Optimum Gamma}: The optimum length of a eukaryotic genome does not depend on the length of the replichore; only on their number. A genome can therefore increase its own length by increasing the length of each of its replichores evenly, without influencing $\gamma$. Such an alteration increases both the replication time and the information storage capacity equally, thereby nullifying its effect on the selection pressure $\gamma$. However, our stochastic model cannot make such concerted changes at multiple locations in the genome, and hence cannot alter the length of the genome this way. When the selection pressure is low, evolution, on the other hand, can alter the genome length by progressively lengthening each of the replichores, by temporarily tolerating an uneven distribution of replichore lengths. This would result in very different genome lengths and corresponding variance in replication times, due to variation in replichore length, even within very similar eukaryotic species, as has been observed abundantly \cite{pagel1992variation, gregory2001coincidence, greenlee1984intraspecific, vsmarda2010understanding}. Therefore, our model does not impose a strict upper limit on the genome length, although it converges to a constant genome length for a fixed number of origins, $Ori_{max}$. The evolutionarily optimized eukaryotic genome length for a constant $Ori_{max}$ value depends on the average length of the initial genome pool, which can be altered to produce a longer or shorter optimized genome.

\subsubsection*{Large variance of eukaryotic genome length due to low selection pressure for replication time}

\begin{figure}[h!]
    \centering
    \includegraphics[width=.8\linewidth]{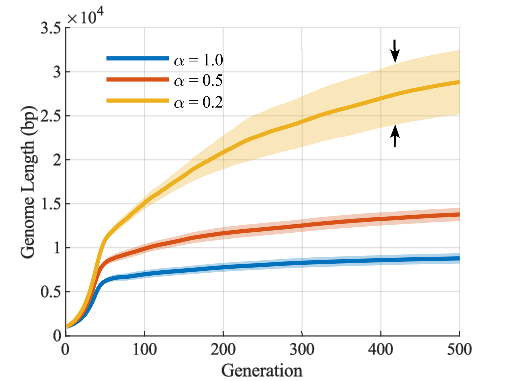}
    \caption{Genome length evolution under reduced selection pressure for replication time. In our model, the replication time is dictated by the length of the longest replichore of the genome. Under low selection pressure for replication time (i.e. low $\alpha$), replichores expand to enhance information storage capacity while minimally impacting the cost of replication time. This drives progressive genome elongation over generations. This reduced pressure also permits uneven replichore lengths across the genome, as opposed to the optimized symmetrical replichore lengths observed under strong selection. Consequently, genome length variability within a set of related species increases over time, as reflected in the widening confidence intervals for low $\alpha$. This explains the genome length variability experimentally observed across species and within conspecific organisms. The confidence interval shown here is calculated with $30$ samples, for the purpose of visual demonstration. The increase in variance with the reduction in $\alpha$ persists for higher sample numbers.}
    \label{fig: Low Selection}
\end{figure}

In order to verify our statement above that the genome length can vary drastically even within similar species when the selection pressure for replication time is low, we reduce the cost of replication time in our evaluation of selection pressure, by raising it to a power $\alpha$, where $0 \le \alpha \le 1$. The modified expression for selection pressure reads
\begin{equation}
    \gamma = \frac{\text{ (replication time)}^{\alpha}}{\text{information storage capacity}}.
\end{equation}
This modification allows for uneven distribution of replichore lengths, and the algorithm tolerates an increase in the replication time by prioritizing information storage capacity, thus enlarging the genome, as explained above. When the importance of replication time reduces, and its cost ($\alpha$) goes down, the variance in the lengths of the genomes of our initial population increases with generation, as seen from the widening of confidence interval with generation in Fig. \ref{fig: Low Selection}.

Such computationally observed significant variation in genome length has been documented across species \cite{bennett2011nuclear, leitch2012genome, gregory2011evolution} and even within conspecific populations \cite{greenlee1984intraspecific, vsmarda2010understanding, mowforth1989intra, lockwood1992genome}, an observation whose evolutionary mechanism is deeply contested. This constitutes the central mystery of the C-value paradox, a set of observations of uncertain evolutionary origins \cite{ohno1972so, gregory2001coincidence}. Prevailing hypotheses posit that persistent upward mutation pressure drives C-value (a measure of genome content) expansion, with species exhibiting slower cellular division rates being more tolerant of random DNA accumulation \cite{pagel1992variation, bennett1972nuclear, petrov2002mutational}. This framework is supported by evidence demonstrating strong negative correlations between genome length and both mitotic and meiotic division rates \cite{vsimova2012geometrical, francis2008strong}. Our computational model provides an explanation for the above observation, where, low selection pressure for replication time shifts the evolutionary trajectory towards maximizing information storage capacity at the expense of replication time. Apart from increasing the genome content over generations (C-value), this also increases the variance of the genome length, leading to drastically different genome lengths even within conspecific organisms (Fig. \ref{fig: Low Selection}). Whether the accumulated genome carries information or not cannot be answered within our model, another mystery of the C-value paradox.   

\section*{Discussion}

The divergence between the lengths of genomes of prokaryotes and eukaryotes has been an enduring enigma, noted since the first systematic C-value measurements carried out in the 1950s. The nearly 100-fold difference in cell sizes between prokaryotes and eukaryotes, and the huge increase in the structural and functional complexity of eukaryotes are generally attributed to this difference. The purpose of this paper is to investigate the evolutionary origin of this divergence. Here, we identify replication time and information storage capacity as the two primary determinants of genome length, with the latter strongly constrained by cellular energetics. The interplay between these two variables dictates the evolutionary trajectory of genome length, for any given species. 

We quantify this interplay between replication time and information storage capacity by defining Selection Pressure simply as the ratio of the above two variables. Evolution acts on the genome, attempting to minimize the replication time while maximizing the information storage capacity. By rewriting the selection pressure in terms of the length of the replichores and the total length of the genome, we make the selection pressure computable. We model the evolution of a pool of genomes by introducing stochastic mutations, randomly adding or deleting a fraction of the total genome, and evaluating the effect of the selection pressure on the mutated genomes. 

We observe that prokaryotic genomes, modeled here as genomes with a \textit{single} replication origin, lose genome length with evolving generations. This is due to the inability of prokaryotic genomes to parallelize their replication by dividing the genome into multiple segments (replichores) that can replicate independently and simultaneously, because of the restriction on the number of origins. This restriction reduces the genome length for prokaryotes, since only two simultaneously replicating segments are allowed, and any increase in the size of these segments increases the replication time, and is thus evolutionarily disfavored. On the other hand, we observe that eukaryotic genomes tend to increase their genome length indefinitely, without incurring a cost from increased replication time, since the model's allowance of a large number of origins for eukaryotic genomes allows for massive parallelization of genome replication, by dividing the genome into multiple independently and simultaneously replicating segments. This indefinite expansion of the eukaryotic genome is curtailed only by cellular energetic constraints, the need for activated monomers and energy supply for simultaneously replicating thousands of genomic segments. This constraint is experimentally demonstrated by the linear relationship between the number of endosymbionts and genome length in eukaryotes (Fig. \ref{fig: MitoVsGenomeSize}). We model this constraint by limiting the number of origins allowed for eukaryotic genomes. 

%%%%%% New %%%%%%%%%
Our model demonstrates that Chargaff’s second parity rule (PR-2), i.e. the symmetry of purine (R) and pyrimidine (Y) frequencies within individual DNA strands, emerges as a direct consequence of selection for replication efficiency. When initialized with purine-filled single-strand sequences, evolution under the defined selection pressures drives them toward parity, yielding strands with equal purine and pyrimidine content. Critically, this symmetry is not a natural outcome of stochastic inter-strand sequence shuffling or thermodynamic entropy maximization, but an adaptive response to selection for balanced replichore lengths. Any compositional bias in a strand would generate asymmetrical replichores, delaying replication and reducing fitness. Thus, PR-2 in our framework reflects an evolutionary optimization: the equalization of R/Y content is a byproduct of selection to harmonize replichore architecture, ensuring efficient bidirectional replication. These results suggest that PR-2 is a signature of replication-driven adaptation, rather than an outcome of neutral processes.

\begin{figure}[h!]
    \centering
    \includegraphics[width=1\linewidth]{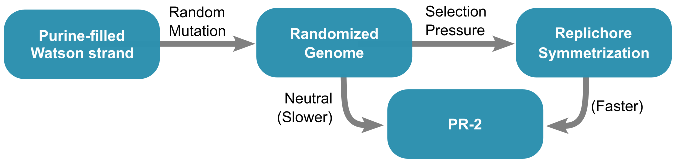}
    \caption{Neutral and adaptive evolutionary trajectories leading to Chargaff's second parity rule (PR-2). In the neutral process, starting from a purine-filled Watson strand of low fitness (high $\gamma$), genomes may reach a state of equal purine (R) and pyrimidine (Y) content on a single strand, i.e. PR-2, solely through a large number of mutations (inversions and inverted transpositions), occurring neutrally without providing any adaptive benefit. The rate of convergence to PR-2 through this route is slower, and the PR-2 compliance is less strict. Moreover, the nucleotide skews that dictate the V-shaped replichore structure are averaged out, nullifying the sequence signature corresponding to replication origins. Alternatively, as demonstrated in this work, selection pressure for rapid replication can \textit{swiftly enforce} PR-2 through replichore length symmetrization, resulting in a \textit{strict} compliance with this parity rule. This adaptive process also retains the sequence signature that determines replication origins.}
    \label{fig: PR2}
\end{figure}

Although not originally intended, surprisingly, our model reproduces several other experimentally observed phenomena, in addition to the prokaryote-eukaryote genome length divergence and reproduction of PR-2 compliance. (a) A general preference for deletion mutations over insertions (Fig. \ref{fig: selection}a) \cite{kuo2009deletional, mira2001deletional, taylor2004occurrence, tao2007patterns, andersson2001pseudogenes, gregory2004insertion}. (b) A tendency to equalize the lengths of all replichores of the genome, as indicated by the optimized $\gamma$ value (Fig. \ref{fig: LenEvo}c,d) \cite{albrecht2007inversions, matthews2010fitness, darling2008dynamics, jespersen2024insertion, prabhu1993symmetry} (c) Correlation between genome length and the number of mitochondria/chloroplasts (Fig. \ref{fig: MitoVsGenomeSize}) \cite{okie2016major, Fahimi2023} (d) Increase in the variance of genome length resulting from a reduced selection pressure for replication time minimization (C-value paradox) (Fig. \ref{fig: Low Selection}) \cite{greenlee1984intraspecific, vsmarda2010understanding, mowforth1989intra, lockwood1992genome} (e) Anticorrelation between cell-cycle time and genome length (Fig. \ref{fig: Low Selection}) \cite{vsimova2012geometrical, francis2008strong}.

Our model can be tested using \textit{in-vitro} evolution experiments on self-replicating DNA sequences that involve deletions and insertion mutations. Under strong selection pressure for faster replication, the sequences should evolve towards a replichore structure with a central origin of replication and equal and opposite nucleotide skews on the two replichore arms. However, when primed to replicate from the ends, these sequences should evolve toward purine/pyrimidine-filled single strands. When the supply of monomers is adequate, the more fit sequences would exhibit multiple origins. Any imbalance in the lengths of the two replichores of the evolutionarily superior sequences should adversely affect the sequence's fitness.

Our explanation for the divergence between prokaryotic and eukaryotic genome lengths rests primarily on the number of origins used during DNA replication: using more origins reduces the replication time, all else being equal. Multicellular eukaryotes appear to have invented a new degree of freedom to modulate their cell replication time depending on tissue-level spatial, temporal, energetic, and environmental constraints by employing an appropriate number of origins. As has been amply demonstrated, multicellular organisms do not utilize all available origins of replication to replicate their DNA. Only about $30\%$ of the origins of the human genome are constitutively fired, with the utilization of the rest depending on the tissue/organism-level requirements \cite{courtot2018protective, sclafani2007cell, aze2018recent}. This top-down control of replication origin firing partly enables these organisms to create specialized organs with disparate cell-cycle rates, such as human skin and colon, where rapid cell-cycle rates are crucial, and neurons in the brain, which rarely replicate, presumably to preserve information \cite{hiratani2008global, ryba2010evolutionarily, gilbert2001making, rhind2013dna}. An important sequence characteristic that segregates multicellular eukaryotic origins into constitutive, latent, and dormant sets is the magnitude of nucleotide skew at the origin locations \cite{guilbaud2022determination}. Our model above too uses these nucleotide skews to identify the locations of origins, although the magnitude of the skew is not utilized for determining the efficiency of origin firing, a simplification that will be removed in a later article. We speculate that the local loss of such top-down control on replication origin selection in various tissues leads to rapid replication and, consequently, carcinogenesis \cite{fu2021dynamics}. Regulation of the number and firing efficiency of replication origins can modulate genome architecture at evolutionary timescales, while organismal top-down control appears to tame the origins into serving the individual organism at the timescale of the lifetime of that individual.

%\nolinenumbers

\section*{Statements and Declarations}

\subsection*{Competing interests}
The authors declare no competing interests.

\subsection*{Acknowledgments}
Support for this work was provided by the Science \& Engineering Research Board (SERB), Department of Science and Technology (DST), India, through a Core Research Grant with file no. CRG/2020/003555 and a MATRICS grant with file no. MTR/2022/000086. 

\subsection*{Supplementary Information}
The algorithms and parameters used to generate the plots are provided in supplementary information.

%%%%%%%%%%%%%%%%%%%%%%%%%%%%%%%%
%%%%%%%%%%%%%%%%%%%%%%%%%%%%%%%%

\section*{Supplementary Information}
\subsection*{Model}

We are interested in the investigation of genome length evolution under a selection pressure that attempts to minimize replication time while maximizing information storage capacity. An initial pool of $N$ identical sequences, composed of all purines or all pyrimidines, are replicated, and the daughter sequences are mutated, producing a pool of $2N$ sequences. Selection acts on this pool, removing $N$ less-fit sequences that do not satisfy the selection pressure adequately. This replication-selection cycle is repeated $m$ times, and the average genome length at every generation is recorded.

\subsection*{Algorithms}

We followed the following algorithm to mutate each sequence in the pool and then calculate the selection pressure $(\gamma)$ to extract the fit sequences.

\subsubsection*{Mutation}
\begin{figure}[h!]
    \centering
    \includegraphics[width=1\linewidth]{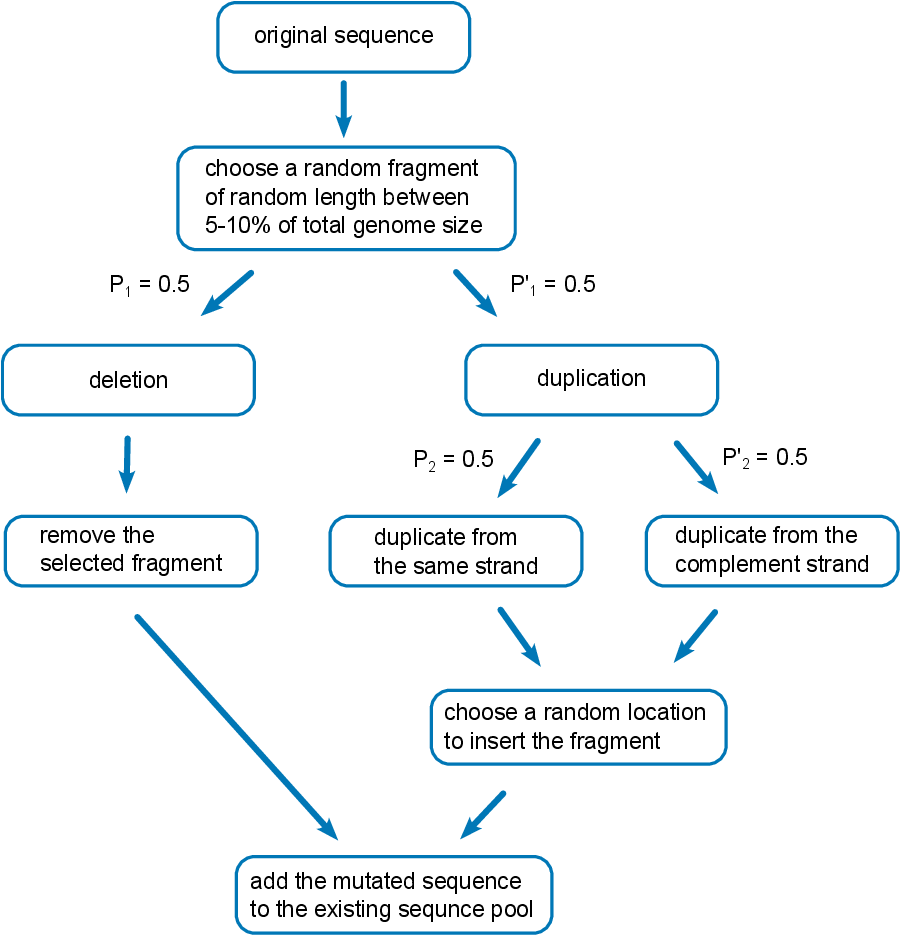}
    \caption{Algorithm for mutation. Here, $P_1$ and $P_1'$ are the probability of the sequence undergoing deletion or duplication, respectively. And, $P_2$ and $P_2'$ are the probability of the sequence undergoing duplication from the same strand or complementary strand, respectively. }
    \label{fig: mutationS}
\end{figure}

We implement large-scale genomic mutations through deletions or duplications of random regions comprising 5\% to 10\% of the total genome length (Fig. \ref{fig: mutationS}). In each generation, every sequence in the pool undergoes a single mutation, i.e., either a duplication or deletion. Following this, the sequence pool is expanded to include the new $N$ mutated sequences along with the $N$ sequences of the previous generation, resulting in a total of $2N$ sequences.

\subsubsection*{Calculation of selection pressure}

Following mutation, $N$ sequences are selected from the pool of $2N$ for the next generation based on a selection pressure aimed at minimizing the factor $\gamma$, which is defined as,
\begin{equation}\label{selection factor modifiedS}
    \gamma = \frac{\textit{length of longest replichore}}{\textit{full genome length}}  
\end{equation}
A low $\gamma$ value of a sequence suggests that the sequence is capable of fast replication and high information storage.

\begin{figure}[h!]
    \centering
    \includegraphics[width=1\linewidth]{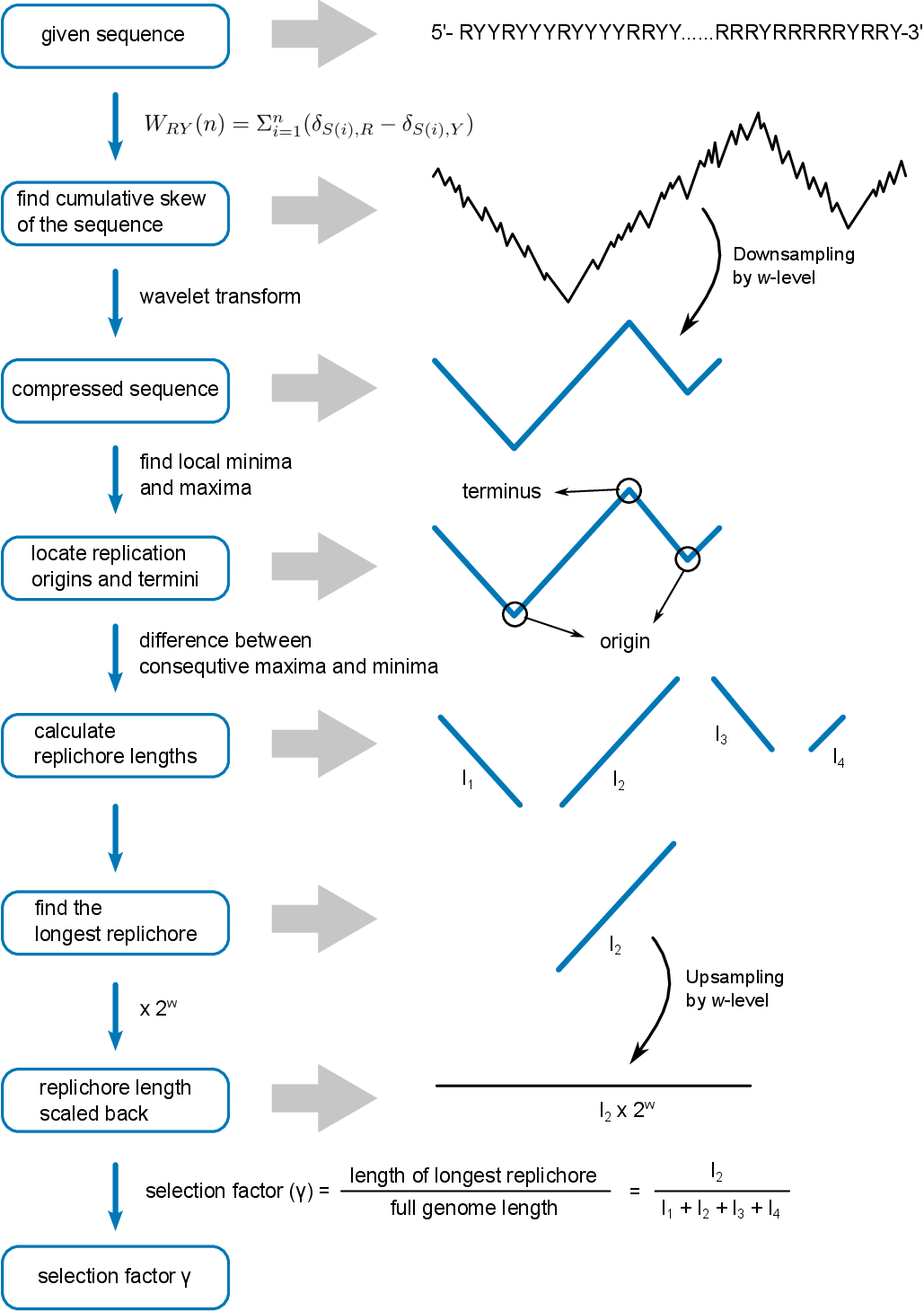}
    \caption{Calculation of selection pressure ($\gamma$)}
    \label{fig: Selection Calc}
\end{figure}
We used cumulative skew of the sequences to identify replication origins and termini and hence the replichore lengths. The purine-pyrimidine (RY) cumulative skew of the sequences, $W_{RY}(n)$, is defined as;
\begin{equation}
    W_{RY}(n) = \Sigma _{i=1}^n (\delta _{S(i),R} - \delta _{S(i),Y})
\end{equation}

Here,$S$ is a genomic sequence of length $N$ bp, composed of four nucleotides, classified into two groups: $R = \{G, A\}$, $Y = \{C, T\}$, and $n = 1 \dots N$.    

The selection pressure $(\gamma)$ is calculated for each of the sequences. Thereafter, $N$ sequences with the lowest $\gamma$ values are selected as evolutionarily superior and are carried forward to the next generation.

\subsection*{Genome Length Evolution Over 1000 Generations}

In our simulations, we set the population size to $N = 1000$, the initial genome length to $1024$ bp ($L_0$), and the number of generations to $m = 1000$, employing a four-level wavelet transformation. Each experiment was repeated 100 times to ensure statistical robustness. The genome length stabilized within 500 generations; thus, for clarity, the main article presents results up to 500 generations. The complete genome length evolution over 1000 generations is shown below.

\begin{figure}[h!]
    \centering
    \includegraphics[width=1\linewidth]{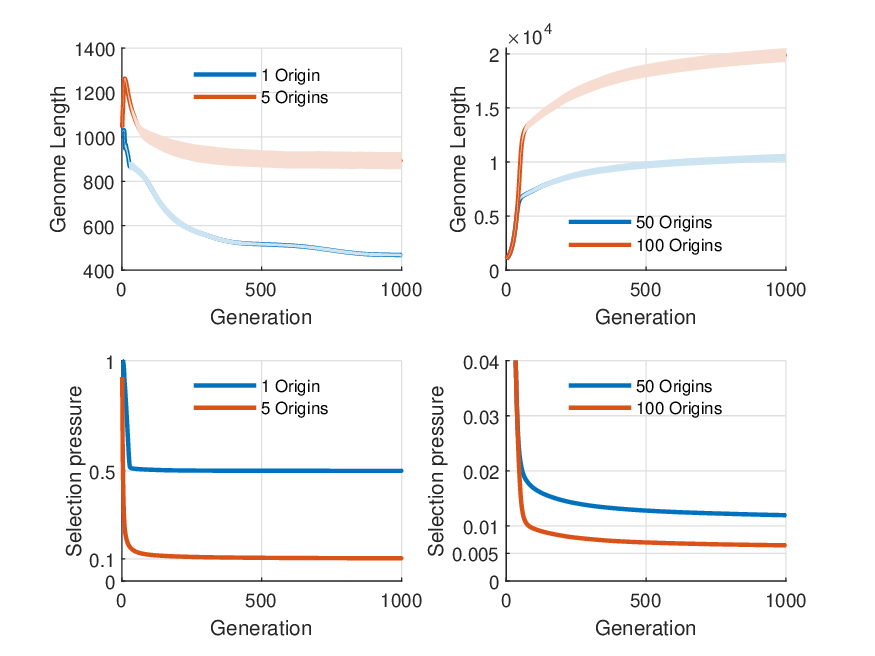}
    \caption{Genome length evolution for 1000 generations}
    \label{fig: 1000Gen}
\end{figure}

\subsection*{Genome Length and Cell Size Allometry}

Genome length has been reported to scale with cell volume following an allometric relationship \cite{gregory2001coincidence, olmo1983nucleotype, mg1973dna}:
\begin{equation}\label{eq: allometry}
V = kC^{\alpha},
\end{equation}
where $V$ represents the cell volume in cubic micrometers, and $C$ denotes genome length in base pairs. We applied this allometric relation to estimate genome lengths from cell volumes in our dataset.

To determine the coefficients $k$ and $\alpha$, we analyzed the correlation between cell volume and genome length for $60$ organisms with available genome data. The estimated coefficients were then used in Eq. \ref{eq: allometry} to infer genome lengths for the remaining organisms in our dataset.

\begin{figure}[h!]
    \centering
    \includegraphics[width=1\linewidth]{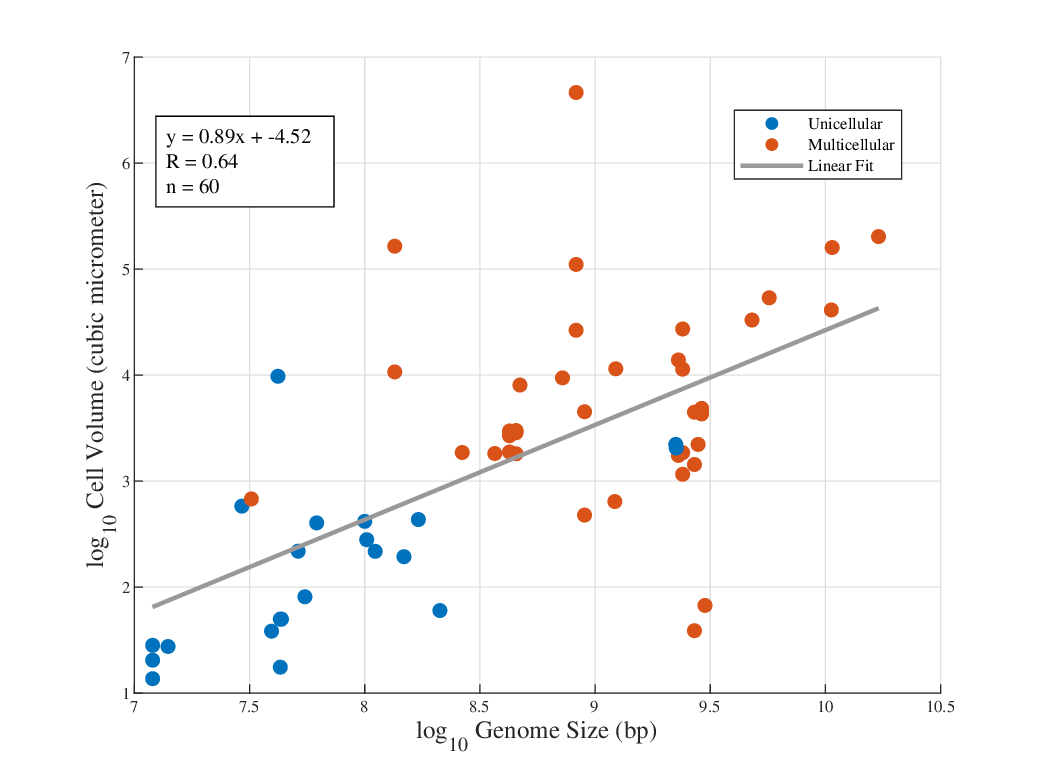}
    \caption{Correlation between genome length and cell volume.}
    \label{fig: correlation}
\end{figure}

The log-log correlation of genome length and cell volume for these $60$ organisms is presented in Fig. \ref{fig: correlation}, yielding estimated coefficients of $k = 3.04 \times 10^{-5}$ $\mu m^3$/bp and $\alpha = 0.89$. Using these coefficients, we estimated genome lengths for the rest of the organisms from equation (\ref{eq: allometry}) and subsequently analyzed their correlation with mitochondria and chloroplast count. The correlation between genome length and mitochondrial count is discussed in the main article.

\subsection*{Parameters}

The following parameters were used to generate Fig. (3) of the main article.
\begin{enumerate}
    \item Length of each sequence in the initial pool, $L_0 = 1024$ bp
    \item Number of sequences in the pool, $N = 1000$
    \item Wavelet level = 4
    \item Minimum length of mutating fragment = $5\%$ of full genome
    \item Maximum length of mutating fragment = $10\%$ of full genome
    \item Number of generations = 1000
\end{enumerate} 

The following parameters were used to generate Fig. (5) of the main article.
\begin{enumerate}
    \item Length of each sequence in the initial pool, $L_0 = 1024$ bp
    \item Number of sequences in the pool, $N = 100$
    \item Wavelet level = 4
    \item Minimum length of mutating fragment = $5\%$ of full genome
    \item Maximum length of mutating fragment = $10\%$ of full genome
    \item Maximum number of replication origins $Ori_{max} = 50$  
    \item Number of generations = 300
\end{enumerate}

The following parameters were used to generate Fig. (7) of the main article.
\begin{enumerate}
    \item Length of each sequence in the initial pool, $L_0 = 1024$ bp
    \item Number of sequences in the pool, $N = 100$
    \item Wavelet level = 4
    \item Minimum length of mutating fragment = $5\%$ of full genome
    \item Maximum length of mutating fragment = $10\%$ of full genome
    \item Maximum number of replication origins $Ori_{max} = 50$  
    \item Number of generations = 500
\end{enumerate}

\printbibliography

\end{document}